\documentclass{optica-article}
\journal{oe}
\articletype{Research Article}
\usepackage{xcolor}
\usepackage{siunitx}

\usepackage{lineno}
\begin{document}
\title{Non-Gaussian quantum state generation by multi-photon subtraction at the telecommunication wavelength}

\author{Mamoru Endo,\authormark{1,2,*} Ruofan He,\authormark{1} Tatsuki Sonoyama,\authormark{1} Kazuma Takahashi,\authormark{1} Takahiro Kashiwazaki,\authormark{3} Takeshi Umeki,\authormark{3} Sachiko Takasu,\authormark{4} Kaori Hattori,\authormark{4,5} Daiji Fukuda,\authormark{4,5} Kosuke Fukui,\authormark{1} Kan Takase,\authormark{1,2} Warit Asavanant,\authormark{1,2}  Petr Marek, \authormark{6} Radim Filip, \authormark{6} and Akira Furusawa\authormark{1,2}}

\address{\authormark{1} Department of Applied Physics, School of Engineering, The University of Tokyo, 7-3-1 Hongo, Bunkyo, Tokyo 113-8656, Japan\\
\authormark{2}Optical Quantum Computing Research Team, RIKEN Center for Quantum Computing, 2-1 Hirosawa, Wako, Saitama 351-0198, Japan\\
\authormark{3}NTT Device Technology Labs., NTT Corporation, 3-1, Morinosato Wakamiya, Atsugi, Kanagawa 243-0198, Japan\\
\authormark{4} National Institute of Advanced Industrial Science and Technology, Tsukuba, Ibaraki 305-8563, Japan\\
\authormark{5}AIST-UTokyo Advanced Operando-Measurement Technology Open Innovation Laboratory, Tsukuba, Ibaraki 305-8563, Japan\\
\authormark{6}Department of Optics, Palacky University,
17. listopadu 1192/12, 77146 Olomouc, Czech Republic
}

\email{\authormark{*}endo@ap.t.u-tokyo.ac.jp} %% email address is required



\begin{abstract}
In the field of continuous-variable quantum information processing, non-Gaussian states with negative values of the Wigner function are crucial for the development of a fault-tolerant universal quantum computer. While several non-Gaussian states have been generated experimentally, none have been created using ultrashort optical wave packets, which are necessary for high-speed quantum computation, in the telecommunication wavelength band where mature optical communication technology is available. In this paper, we present the generation of non-Gaussian states on wave packets with a short 8-ps duration in the \SI{1545.32}{nm} telecommunication wavelength band using photon subtraction up to three photons. We used a low-loss, quasi-single spatial mode waveguide optical parametric amplifier, a superconducting transition edge sensor, and a phase-locked pulsed homodyne measurement system to observe negative values of the Wigner function without loss correction up to three-photon subtraction. These results can be extended to the generation of more complicated non-Gaussian states and are a key technology in the pursuit of high-speed optical quantum computation. 
\end{abstract}

%%%%%%%%%%%%%%%%%%%%%%%%%%  body  %%%%%%%%%%%%%%%%%%%%%%%%%%
\section{Introduction}
Continuous-variable quantum information processing (CVQIP), in which quantum information is encoded in the quadrature-phase amplitude of the electromagnetic field of light, has the potential to enable the creation of ultra-fast, fault-tolerant quantum computers \cite{Takeda2019}. This form of processing has a high carrier frequency, which can increase the upper limit of clock frequency, and has low decoherence even at room temperature and atmospheric pressure due to the high photon energy compared to the environment. CVQIP is also compatible with measurement-based quantum computation (MBQC), which allows for programmable quantum computation \cite{Ukai2015}. In MBQC, a multipartite quantum entangled state called a cluster state is prepared and local measurements of qubits are performed. By making the appropriate measurements, the quantum computation can be performed by reading out the final quantum state.

Using a traveling wave of light, cluster states can be deterministically constructed using squeezed light sources \cite{Andersen2016} and asymmetric interferometers. The measurements are made with homodyne detectors, and various measurements can be made by choosing the measuring phase of local oscillator (LO) lights. This cluster state, called a time-domain multiplexed (TDM) cluster state, has already been realized on a large scale \cite{Yokoyama2013,Yoshikawa2016,Asavanant2019,Larsen2019}, and basic quantum operations using it have also been demonstrated \cite{Asavanant2021,Larsen2021}. In recent years, there have been significant developments in squeezed light sources \cite{Kashiwazaki2021,Nehra2022}, and homodyne detectors \cite{Tasker2020,Bruynsteen2021,Inoue2022}. In particular, squeezed light with a squeezing level and bandwidth of \SI{6}{dB} and \SI{6}{THz} can be easily generated from a waveguide optical parametric amplifier (WG-OPA) in the telecommunication wavelength band around \SI{1545}{nm} \cite{Kashiwazaki2021}. Broadband homodyne detection with a bandwidth of over \SI{40}{GHz} has also been achieved using technologies developed for optical telecommunications \cite{Inoue2022}. This means that quantum information can be encoded in wave packets with picosecond-order pulse duration, indicating that high-clock frequency quantum computation may be possible even with current technology.

However, the TDM cluster state and homodyne detectors alone are not sufficient for constructing a fault-tolerant universal quantum computer. In order to achieve fault tolerance, quantum states called non-Gaussian states, such as Gottesman-Kitav-Preskill (GKP) qubits \cite{Gottesman2001}, must be introduced to the cluster state \cite{Alexander2018,Niklas2022}. The generation of non-Gaussian states is one of the major challenges in CVQIP.

Previously, many experiments generating non-Gaussian states with strong Wigner negativity have been conducted using continuous-wave (cw) lasers in the near-infrared region and relatively long wave packet sizes (from 10 to \SI{100}{nm} range) \cite{Neergaard-Nielsen2006,Neergaard-Nielsen2007,Wakui2007,Takahashi2008,Yukawa2013,Asavanant2017,Bouillard2019}. These experiments have been successful due to the availability of high-performance photon detectors and the ability to perform high-precision quantum state tomography using high-efficiency homodyne measurements. However, for practical CVQIP applications, it is desirable to use the telecommunication wavelength band, where broadband squeezed light sources, homodyne detectors, and various other high-quality optical components are available.
Additionally, the wave packets of non-Gaussian states should be shortened in order to take advantage of the potential of CVQIP for ultrafast quantum calculations. Furthermore, considering the need for photon-number resolving detection and the timing synchronization with TDM cluster states, the use of pulsed light sources is preferable. 
Therefore, it is important to realize non-Gaussian states in the telecommunication wavelength band and with pulsed light sources in order to move beyond proof-of-principle demonstrations and achieve actual optical quantum computation.

Recent advances in superconducting photon detectors at telecommunication wavelengths \cite{Lita2008,Miki2017,Reddy2020,Lita2022,Hattori2022} have enabled the observation of non-Gaussian states with Wigner negativity in the continuous-wave (cw) regime \cite{Kawasaki2022,Takase2022}. Additionally, wavelength conversion methods have been used to generate non-Gaussian states \cite{Baune2017}. However, negative values have not yet been obtained in experiments using pulsed light sources in this wavelength \cite{Namekata2010}.

Experiments with pulsed light are more challenging than those with cw light due to the difficulty in matching spatial and temporal modes in pulse homodyne measurements. Therefore, there are only a few cases in which experiments with pulsed lasers have been able to observe negative values of the Wigner function without loss correction \cite{Cooper2013,Gerrits2010,Ra2019}.

In order to accurately reconstruct the Wigner function through quantum state tomography, the phase of the homodyne measurements must be accurately controlled. Most previous experiments with pulse light have performed quantum state tomography by scanning the local oscillator (LO) phase without fixing it, resulting in an insufficient number of data points per phase and inaccurate quantum state estimation due to phase oscillation effects. For applications in optical quantum computing, phase-locked single-shot measurements must be necessary. 

In this paper, we generated pulsed squeezed light in the telecommunication wavelength band (\SI{1545.32}{nm}) using a WG-OPA module with a quasi-single spatial mode \cite{Kashiwazaki2021}. Non-Gaussian states, known as Schr\"{o}dinger cat states, were generated through the photon subtraction method \cite{Danka1997}, in which a small portion of the squeezed light is detected by a photon-number resolving detector (PNRD) based on a superconducting transition edge sensor (TES). Our low-loss quasi-single spatial mode WG-OPA, high-performance PNRD, and waveform shaping of the phase-controlled LO light successfully produced a negative value of the Wigner function without any loss correction, a first for a pulsed light source in the telecommunication wavelength band to the best of our knowledge. The wave packet of the generated state had a width of approximately \SI{8}{ps}, which is compatible with broadband TDM cluster state resources and homodyne detectors. These results demonstrate that the presented non-Gaussian state generation system is suitable for use in CVQIP and is a significant step towards realizing an ultra-fast optical quantum computer with fault-tolerance.

\begin{figure}[htbp]
  \centering
  \includegraphics[width=13cm]{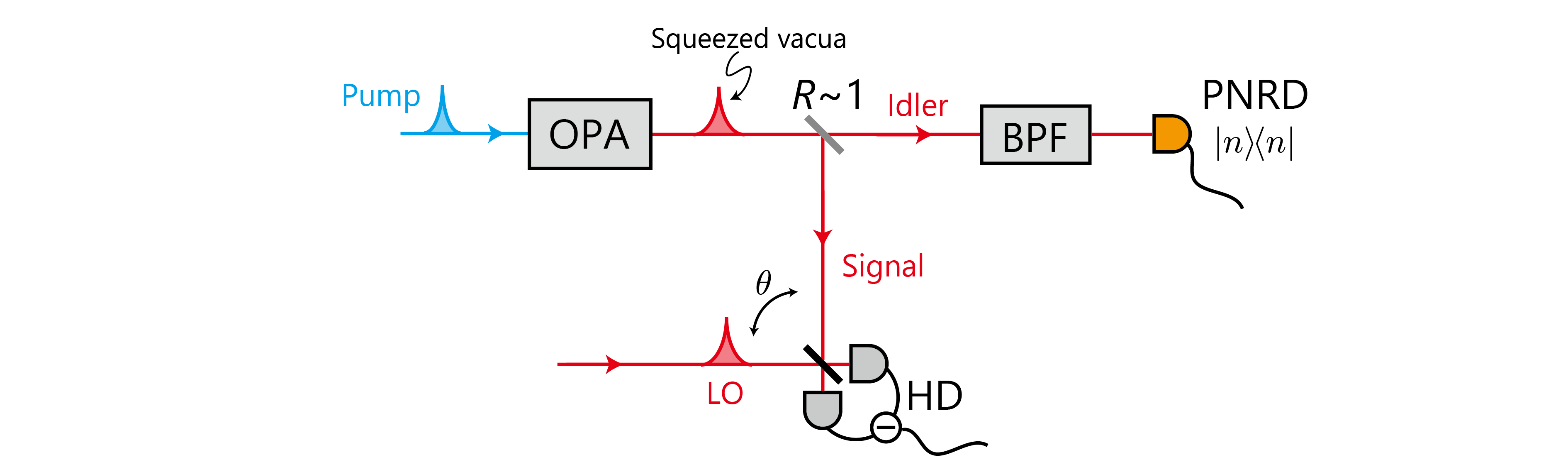}
\caption{The conceptual image of photon subtraction experiment. An OPA pumped by a pulsed laser emits squeezed vacua. A small portion of this is picked up by a beamsplitter with a refrectance of $R$ and detected by a photon number resolving detector (PNRD). An optical bandpass filter (BPF) is typically used for removing unwanted wavelength components. The rest is measured by a homodyne detector (HD). The phase $\theta$ represents measurement phase of the HD.}\label{fig.1}
\end{figure}

\section{Experiment}
In this section, we explain the concept of photon subtraction followed by experimental details. 
\subsection{Concept of photon subtraction}
Squeezed vacuum or squeezed light can be generated using an optical parametric amplifier. Squeezed light is a Gaussian state, so the Wigner function has no negative values. When a small portion of the output is split by a beamsplitter with a reflectance of $R$ (typically $R\sim 1$), as shown in Fig. \ref{fig.1}, the two outputs become quantum entangled. These two outputs are called the signal and idler. The idler is then measured by a PNRD (or just a single-photon detector). Photon-number resolving detection is non-classical and the non-classicality propagates to the signal through quantum entanglement, resulting in a state with negative values of the Wigner function on the signal. This method, in which the PNRD appears to subtract photons from the squeezed light, is called photon subtraction and is one of the heralded methods \cite{Danka1997,Neergaard-Nielsen2006,Wakui2007,Takase2022,Takase2022-2}. The state generated in this way is called a photon-subtracted squeezed state and is known to be a good approximation of Schr\"{o}dinger cat states or coherent-state superpositions, which are important states that can be used to generate the most promising error-correcting code such as GKP qubits \cite{Gottesman2001}.

To evaluate the generated states, quantum state tomography is performed \cite{Lvovsky2009}. The quadrature phase amplitude given by $\hat{x}_\theta=\frac{\hat{a}e^{-i\theta}+\hat{a}^\dagger e^{i\theta}}{\sqrt{2}}$ is measured by a balanced homodyne detector (HD) as shown in Fig.~\ref{fig.1}, where $\hat{a}, \hat{a}^\dagger$ are annihilation and creation operators, which satisfy $[\hat{a},\hat{a}^\dagger]=1$, and $\hbar=1$.  The phase of the local oscillator (LO) $\theta$ can be controlled by a phase modulator. 

Negative values of the Wigner function are easily lost due to experimental imperfections (loss or phase fluctuations) and cannot be recovered through classical operations. In other words, the fact that a state with a negative value of the Wigner function is obtained without loss correction indicates that the experiment is performed accurately with sufficiently small loss.

\subsection{Experimental apparatus}
\begin{figure}[htbp]
  \centering
  \includegraphics[width=13cm]{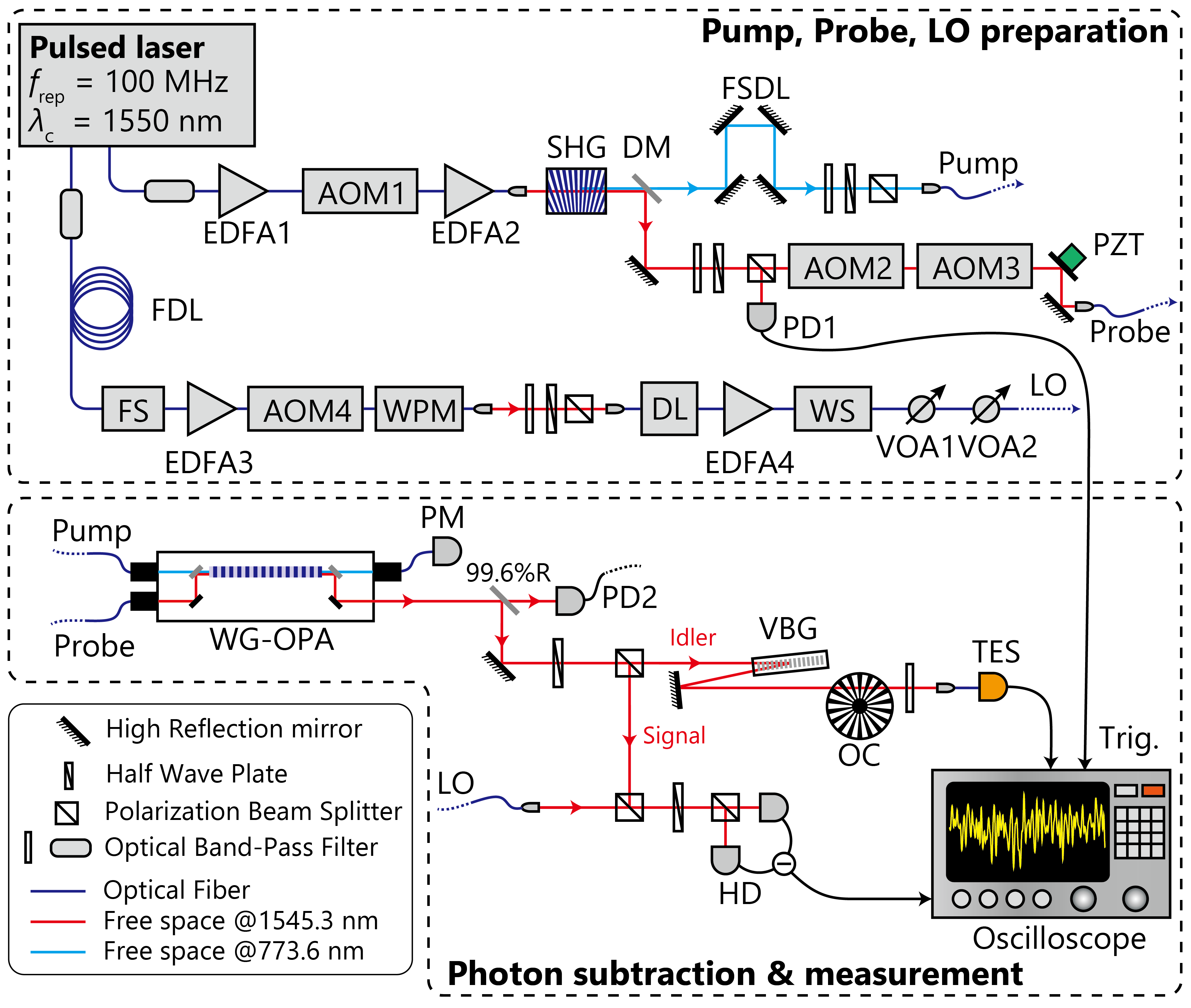}
\caption{(a) The experimental apparatus. EDFAs, erbium-doped fiber amplifiers; AOMs, acousto-optic modulators; SHG, second-harmonic generation crystal; DM, dichroic mirror; FSDL, free-space delay line; PDs, photodiodes; PZT, piezoelectric transducer; FDL, fiber delay line; FS, fiber stretcher; WS, WaveShaper;  WG-OPA, waveguide optical parametric amplifier; PM, power meter; VBG, volume-Bragg grating; OC, optical chopper; TES, transition-edge sensor; LO, local oscillator light; HD, homodyne detector, Trig.; trigger for the oscilloscope. }\label{fig.2}
\end{figure}

\subsubsection{Preparation of pump, probe and LO lights}
Figure \ref{fig.2} shows the experimental apparatus. 
The original pulsed laser source was a femtosecond mode-locked fiber laser at a center wavelength of \SI{1550}{nm} with a spectrum width of more than \SI{40}{nm}, and a repetition frequency of \SI{100}{MHz} (M-Comb-ULN, Menlo Systems). 
Due to the timing resolution limitation of the TES (approximately \SI{50}{nm}) used in this experiment, the repetition frequency of the pump light for WG-OPA was lowered to \SI{5}{MHz} by pulse picking with AOMs, and the broadband spectrum was limited by bandpass filters to create a non-Gaussian state with a wave packet width of about \SI{8}{ps}. The probe light used to reference the phase of the squeezed light via parametric process, and the LO used for homodyne measurements were also pulse-picked in the same way to limit the spectrum. The details are explained in the followings.

The mode-locked laser has multiple output ports, one of which was used to create the pump and probe lights. Note that all of optical fibers used in this experiment were polarization maintaining single-mode fibers. The spectrum of the output light was first limited to a center wavelength of \SI{1545.32}{nm} and a spectral width of \SI{1.0}{nm} by an optical bandpass filter, and then amplified by a home-made erbium-doped fiber amplifier (EDFA1) with a gain fiber (Er80-8/125-PM, nLight). A fiber-coupled acouto-optic modulator (AOM1; SGTF400-1550-1P(H), Chongqing Smart Science \& Technology Development) with a driving frequency of \SI{400}{MHz} was used for pulse picking, and the repetition rate was reduced to \SI{5}{MHz}. The driving and gating signals for pulse picking were derived from the repetition frequency of the mode-locked laser \cite{de-Vries2015}.  The pulse was amplified again by a second EDFA (EDFA2) and emitted into free space. An erbium/ytterbium co-doped double-clad fiber with a large core diameter (DCF-EY-10/128-PM, CorActive) was used for the gain fiber of EDFA2 to prevent pulse distortion due to nonlinear effects caused by high peak power. The light was focused onto a fan-out periodically-polled MgO-doped near-stoichiometric $\text{LiTaO}_3$ crystal (fan-out PPMgSLT, OXIDE) as a second harmonic generation crystal (SHG). The second harmonic light had a center wavelength of \SI{772.66}{nm}, a spectral width of \SI{0.5}{nm}. The second harmonic and fundamental lights were bifurcated by a dichroic mirror (DM) and used as pump and probe lights, respectively. After free-space optical delay line (FSDL) for timing adjustment between pump and probe lights, the pump light was power-adjusted by a half-wave plate (HWP) and a polarization beamsplitter (PBS), and then injected into a polarization-maintaining single-mode fiber. Part of the fundamental light was split by a HWP and a PBS, and was received by PD1 and used to trigger the measurement. The rest of the fundamental light was used as a probe light. The fundamental light was chopped at a frequency of \SI{200}{Hz} by two free-space AOMs (AOM2 and AOM3; AOM3080-1912, Gooch \& Housego) for use in phase-locking by the sample-and-hold method. In addition, to perform heterodyne beat locking in this experiment, the two AOMs were driven at frequencies of \SI{-80.0}{MHz} and \SI{80.1}{MHz}, respectively, with a detuning of \SI{100}{kHz}. A mirror with piezo (PZT) was placed in the optical path for phase locking, and coupled to a polarization-maintaining single-mode fiber.

A LO light was made from another output port of the mode-locked laser. Most prior experiments using pulsed light have used light emitted from a single output as the LO light and pump light \cite{Ourjoumtsev2007,Gerrits2010,Namekata2010,Bouillard2019}. In this case, the pulsed squeezed light has a shorter pulse duration and wider spectral width than the LO light \cite{Gerrits2010,Eto2011}. In addition, it is affected by dispersions in the SHG and OPA \cite{Taguchi2020}, so it is challenging to obtain high squeezing level and large Wigner negativity without pulse shaping \cite{Gerrits2010,Eto2011}. In our experiment, we used light from an output port different from that used to generate the pump and probe lights, and use a commercial waveform shaper (WS; WaveShaper 1000A, II-VI) to achieve more flexible LO light waveform shaping. The output spectrum was limited by a bandpass filter with a center wavelength of \SI{1545.32}{nm} and a spectral width of \SI{12}{nm}. Pulses were pre-chirped (stretched) by a 32-m fiber delay line (FDL) to prevent the system from being distorted by amplification by EDFAs in the latter stage. For long-term feedback control of phase stabilization, a fiber stretcher (FS) with a cylindrical piezoelectric transducer (PT140.70, PI) was included. After amplification by EDFA3 (EDFA100P, Thorlabs), the repetition frequency was reduced to \SI{5}{MHz} by pulse picking with a fiber-coupled AOM (AOM4; SGTF400-1550-1P(H), Chongqing Smart Science \& Technology Development). It was further passed through a waveguide phase modulator (WPM; MPX-LN-0.1-00-P-P-FA-FA, iXblue) for feedback control. After passing through a bandpass filter to cut the amplified spontaneous emission generated by EDFA3, a HWP and a PBS were used for power adjustment. The light was coupled to a polarization-maintaining fiber, and the optical path length was fine-tuned through an optical delay line (DL; ODL-600-11-1550-8/125-P-60-3A3A-1-1-600, OZ optics). The light was amplified again by EDFA4 (EDFA100P, Thorlabs). The spectral shape and phase were then adjusted by the WS. After passing through a variable optical attenuator (VOA1 and VOA2; MMVOA-1-1550-P-8/125-3A3A-1-0.5, OZ optics) used for power stabilization and power modulation of the LO light described later, the light was outgoing to free space with a collimator and used as the LO light. In our experiment, WS was set to be Gaussian with a center wavelength of \SI{1545.32}{nm} and a spectral width of \SI{0.8}{nm}. For the spectral phase, a second-order dispersion of \SI{-1.0}{ps/nm} was added to compensate for the dispersion added by the optical elements. The pulse duration measured by an autocorrelator was \SI{8}{nm}, which corresponds to the trans-form limited pulse duration.

\subsubsection{Photon subtraction and measurement}
A squeezed light was generated by pumping a WG-OPA by the pump pulse with a pulse energy of \SI{0.8}{pJ}. The pulse energy was always monitored by a power meter (PM). The WG-OPA was the same as our previous researches, where we generated non-Gaussian state with cw light source \cite{Takase2022,Takase2022-2}, and the details of the WG-OPA can be found in the reference \cite{Kashiwazaki2021}.
To refer the phase of squeezed light, the probe light was also introduced to the WG-OPA, and a small portion (0.4\%) of the output was detected by a photodetector (PD2), which measures a parametric gain due to the pump light. Because the optical frequency of the probe was slightly shifted by AOM2 and AOM3, the beat signal of the frequency of 200 kHz was obtained, and this beat signal was used for lock-in detection to generate an error signal. The phase between the pump and probe, that is, the phase between the squeezed light and the probe light was locked by a mirror with the PZT.
The squeezed light is then split into signal and idler by a PBS . A HWP before the PBS controls the splitting ratio. In this paper, the reflectance was set to 97\%, 92.4\%, and 88.3\% for one-, two-, and three-photon subtraction, respectively. For zero-photon subtraction experiment, the reflectance was set to 97\%.
The idler light was passed through a volume Bragg grating (VBG; SPC-1544, OptiGrate) and a dielectric multi-layer filter as optical bandpass filters to remove unwanted wavelength components, including the pump lights for the WG-OPA (\SI{772.66}{nm}) and for EDFAs (\SI{976}{nm}), and amplified spontaneous emission of EDFAs. 
The total bandwidth and efficiency of these filters were \SI{0.3}{nm} at the center wavelength of \SI{1545.32}{nm}, and 98\%, respectively. 
An optical chopper (OC; MC1000A, Thorlabs) was used to block the probe light during the measurement. 
The idler light was coupled to a single-mode fiber and detected by an Au/Ti bilayer TES cooled down to \SI{250}{mK} by an adiabatic demagnetization refrigerator (ADR-M, Entropy).  The TES acted as a PNRD and could count more than ten photons at the wavelength of \SI{1545}{nm} with a detection efficiency of approximately 70\%. 
The signal of the TES was amplified by a superconducting quantum interference device (SQUID) array (C6X16FA, Magnicon) and a SQUID controller (XXF-1-20, Magnicon).   
The total efficiency of the idler path was 60\% and a dark count was negligible.
The signal light was measured by a homodyne detector (HD).
A pulse energy of LO was set to be \SI{16}{pJ} and the average power of LO was stabilized by feedback control with VOA1.  
Two PBSs and one HWP compose 50\% beamsplitter and the two outputs were detected by a home-made balanced photodetector with two InGaAs photodiodes.
Quantum efficiency of each photodiode was higher than 96\% at the wavelength of \SI{1545}{nm}. 
The losses for each part of this experiment are listed in Table \ref{tab.loss}. 

\begin{table}[ht]
\centering
\caption{Loss budget in the signal mode}
\label{tab.loss}
\begin{tabular}{lc}
\hline
Optical element & Loss\\
\hline
WG-OPA & 11\% (estimated) \\
Inefficiency of photodiodes for HD & 4\% \\
Spatial mode matching & 4\% \\
Temporal mode matching &  12\% (estimated)\\
Propagation loss & 5\% \\
Circuit noise & 5\%\\
\hline
\end{tabular}
\end{table}

\begin{figure}[htbp]
  \centering
  \includegraphics[width=13cm]{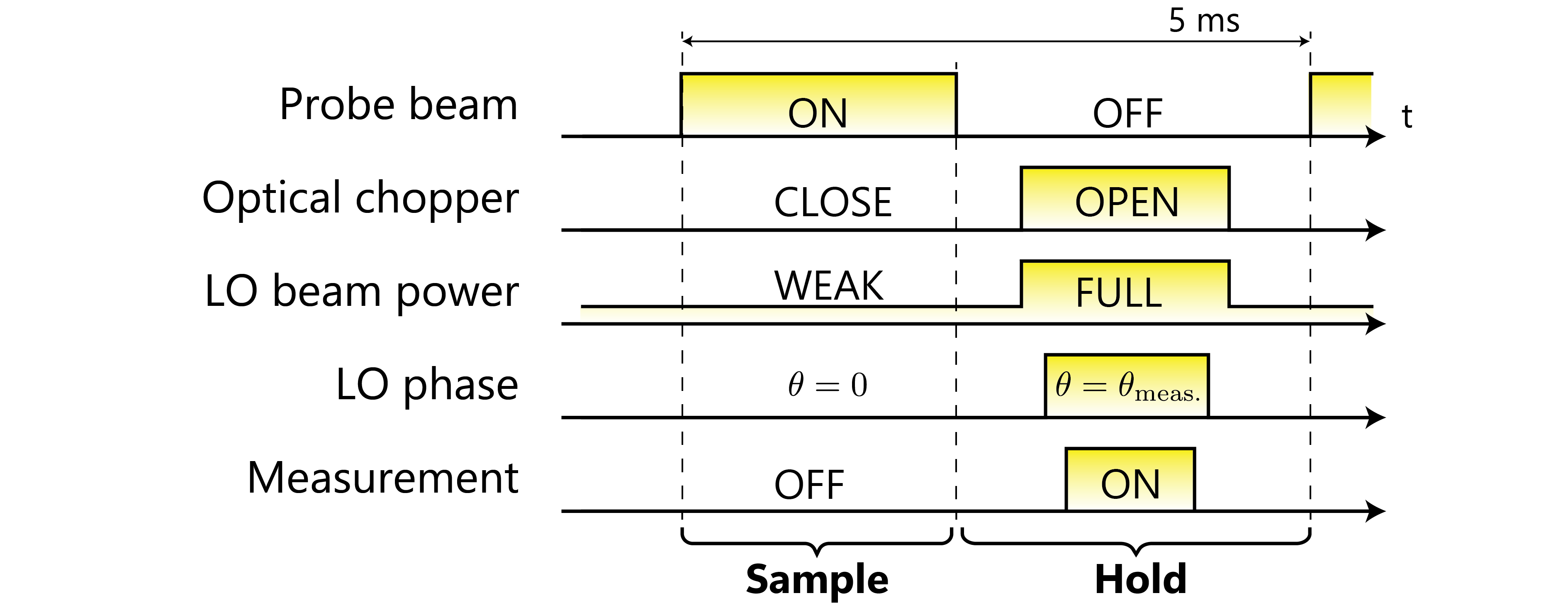}
\caption{The measurement sequence.}\label{fig.3}
\end{figure}

The phase of homodyne measurements $\theta$ was monitored via a beat signal between LO and probe lights at a frequency of \SI{100}{kHz}.
This phase is locked by WPM in the LO path.

The probe light was used to lock the phase of the squeezed light and the measurement phase of the homodyne measurement. If the probe light is present during the measurement, the quantum state is spoiled. Therefore, the sample-and-hold method was applied, which is often used in non-Gaussian state generation experiments with cw light \cite{Asavanant2017,Nehra2019,Takase2022,Takase2022-2}. During the sample phase, the probe light was turned on to measure and stabilize the phases of the experimental system by feedback control. During the hold phase, the probe light was turned off and the system maintain the sample state, and the measurement was performed. The sample-and-hold sequence is shown in Fig. \ref{fig.3}.

As mentioned above, the probe light was chopped by two AOMs with a frequency of \SI{200}{Hz} (5-ms interval). The OC on the idler side was set to open when the probe light was off. When locking the phase of homodyne measurement, the output of the HD was saturated because the LO light power was too strong, and the phase could not be measured. Therefore, VOA2 on the LO side attenuates the LO optical power by about 20 dB during the sample phase. The phase of the homodyne measurement was always locked to a certain fixed value during sample (e.g. $\theta=0$), and the measurement phase of the homodyne measurement was changed to an arbitrary value by applying a voltage to WPM during the hold phase. WPM was the same as that used for feedback control of the phase of the homodyne measurement, and the feedback signal and the signal for phase adjustment were added together by an summation circuit.
The signal from the TES was analyzed when the optical chopper was open, and the waveform from the homodyne detector was acquired by an oscilloscope (MSO64B, Tektronix) when the desired number of photons was detected. Considering the time required for the optical chopper to be completely open, the data acquisition time was set to approximately \SI{0.75}{ms} out of the hold time. This means that the duty ratio of the measurement was 15\%.

When the desired number of photons was detected by the TES, the output voltage signal of the HD was recorded. Acquired 10,000 (zero- and one-photon subtraction) or 5,000 (two- and three-photon subtraction) frames for one LO phase and measure in seven bases ($\theta = 0^\circ, 15^\circ, 30^\circ, 45^\circ, 60^\circ, 75^\circ$, and $90^\circ$). From the obtained data set of quadrature phase amplitudes, the density matrix of the quantum state was estimated by the maximum likelihood estimation method \cite{Lvovsky2009}, and the Wigner function was calculated from the density matrix. Note that the measurement was made with the signal side blocked as the shot noise of the vacuum field, which was used for normalization.

The count rates of this experiment are approximately 3,000 /s, 200 /s, and 5 /s for one-, two-, and three-photon subtraction, where the measurement duty ratio of 15\% is not taken into account. The data acquisition time of this measurement was about three hours for three-photon subtraction, and the phase locking, power fluctuation, and alignment of each part during the measurement were sufficiently stable. In fact, the locks did not drop for more than several days under normal laboratory conditions, and the system can be applied to experiments that require much longer measurement time. In more complicated non-Gaussian state generation, it is necessary to interfere multiple squeezed lights with beam splitters or to interfere with coherent lights called displacement beams \cite{Yukawa2013,Su2019,Asavanant2021-2,Takase2021,Fukui2022,Takase2022-3}. The proposed method can be easily extended to these cases. Note that phase fluctuation effects will occur for disturbances with frequency components faster than the sample-and-hold frequency of \SI{200}{Hz}. The sample-and-hold frequency in this experiment is limited by the bandwidth of VOA2 used for LO light intensity modulation. By using a VOA with wider bandwidth, it is possible to increase the sample-and-hold frequency and realize a robust experimental system against external disturbances.

\section{Result and discussion}

\begin{figure}[htbp]
  \centering
  \includegraphics[width=13cm]{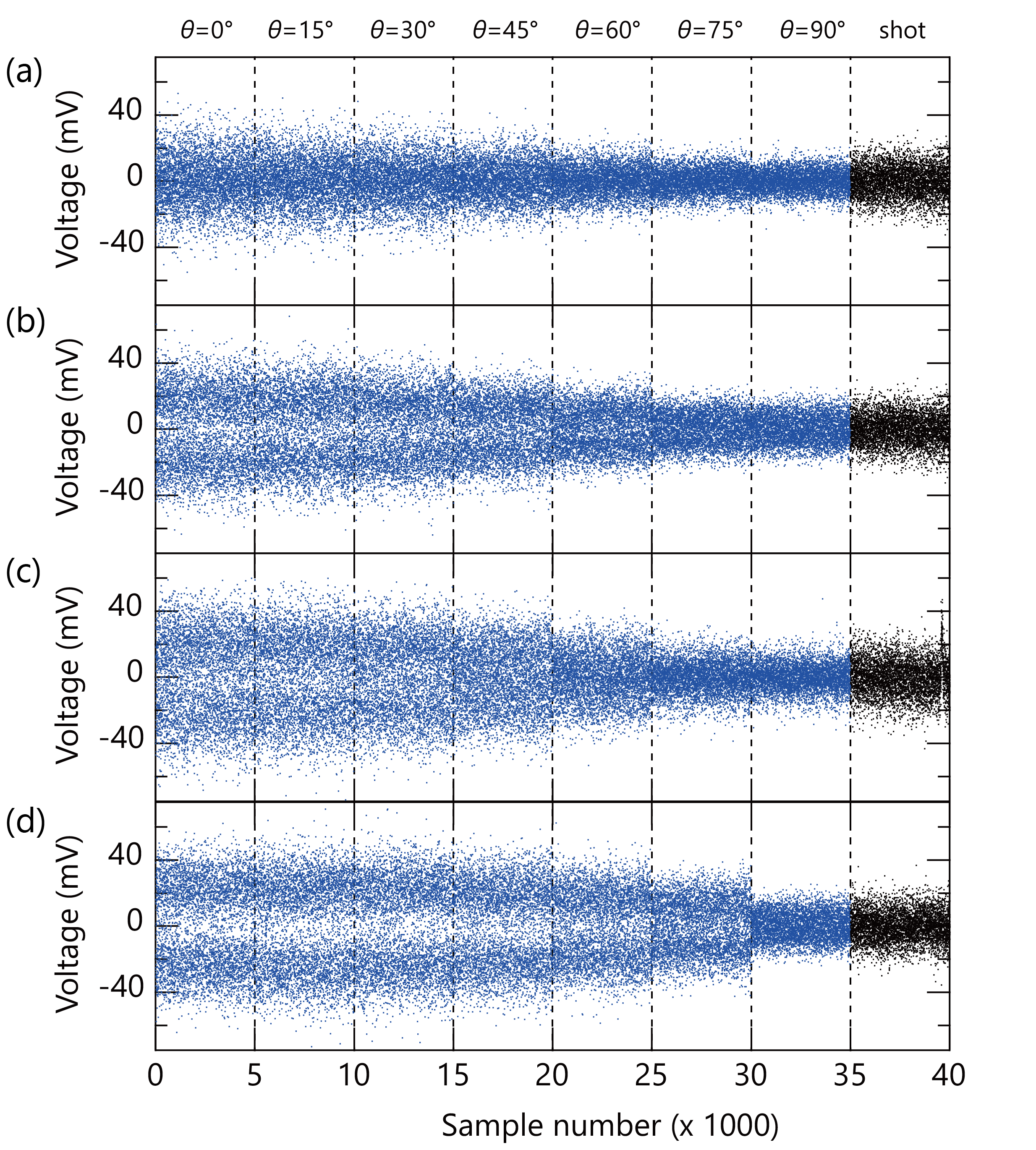}
\caption{The measured voltage signal of the HD, corresponds to quadrature phase amplitude, for each LO phase ($\theta=0,15,30,45,60,75,$ and $90^\circ$), when the subtraction photon number is (a) zero, (b) one, (c) two, and (d) three. The shot noise data for calibration are also shown.}\label{fig.4}
\end{figure}
\begin{figure}[htbp]
  \centering
  \includegraphics[width=13cm]{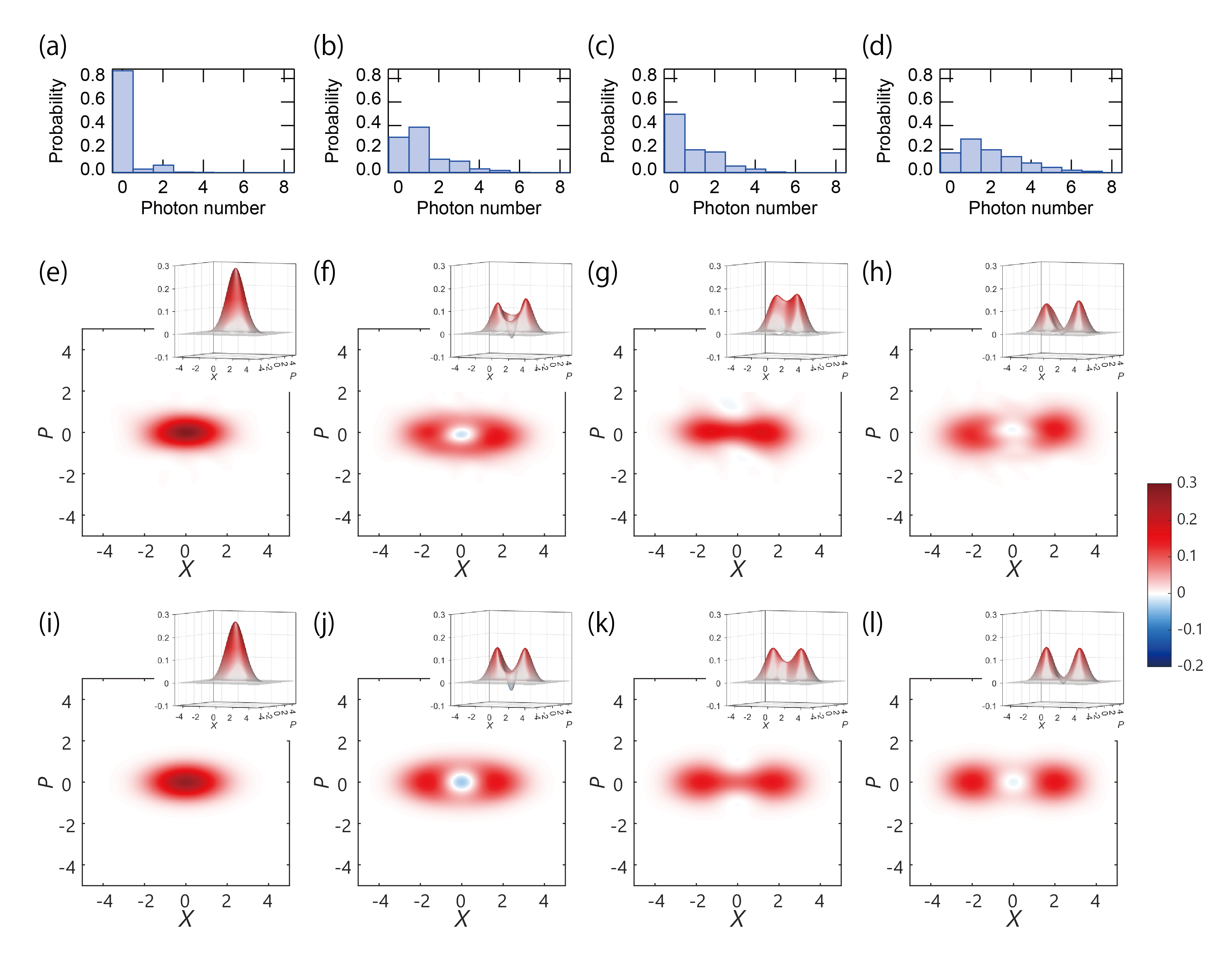}
\caption{(a)-(h) The results of quantum state tomography without loss correction, when the subtraction photon number is (a)(e) zero, (b)(f) one, (c)(g) two, and (d)(h) three. (a)-(d) The photon number distributions. (e)-(h) Reconstracted Wigner functions with a colorbar. (i)-(l) Simulated Wigner functions.}\label{fig.5}
\end{figure}

Figure \ref{fig.4} shows the output voltage signal of the HD, which corresponds to the measured values of quadrature phase amplitude, with different LO phases ($\theta = 0^\circ, 15^\circ, 30^\circ, 45^\circ, 60^\circ, 75^\circ$, and $90^\circ$), when the TES detects photon(s). The shot noise data is also shown in black dots. From Fig.~\ref{fig.4}(a), the initial squeezing and anti-squeezing levels are estimated to \SI{2.9}{dB} and \SI{4.4}{dB}, respectively. It can also be seen that when the TES subtracts photon(s), the distribution has a low density near the origin. Note that, Fig.~\ref{fig.4}(a) and (b) plot first 5,000 points out of 10,000 for comparison. 
Figure \ref{fig.5} (a)-(h) shows the results of quantum state tomography when the TES detects zero, one, two, and three photons. Figure \ref{fig.5}(i)-(l) show simulated Wigner functions based on our loss estimation (Tab. \ref{tab.loss}) using Strawberry Fields \cite{Killoran2019,Bromley2020}.
The photon number distributions and Wigner functions calculated from the density matrices are shown in Fig. \ref{fig.5}(a)-(d) and (e)-(h), respectively. Note that, in this paper, we do not correct any kind of losses. The photon number distributions show that the odd components are increased by subtracting one and three photons from the squeezed light, which is an even photon stream. 
The Wigner function also clearly shows that there is a negative value region near the origin ($W(0,0) = -0.020\pm0.005$ and $W(0,0) = -0.0047\pm0.002$, when one- and three-photon subtraction, respectively). 
In the case of two-photon subtraction, negative values do not appear at the origin. From the simulation result, the minimum value of the Wigner function is $-0.0030$ at the points of $(X,P)=(0,-1.12),(0,1.12)$ (Fig. \ref{fig.5}(k)). In the experimental result (Fig. \ref{fig.5}(g)), the values of reconstructed Wigner function at the same points are $-0.0025\pm{0.001}$ and $-0.0027\pm{0.001}$, respectively, which is in good agreement with the simulation.
Also, the nonclassical effect also narrows the phase space sub-Planck structure \cite{Zurek2001} of the Wigner function in the momentum quadrature ($P$), similarly as for the GKP qubits. We calculate phase-space variance after compensating for the negativity by using a normalized version of positive $|W(0,p)| + W(0,p)$ as a witness for this sub-Planck structure. Thus we compare only the positive parts of functions $W(0,p)$ from figures (Fig. \ref{fig.5}(g) and Fig. \ref{fig.5}(e)) and observe roughly $20\%$ narrowing of the sub-Planck structure, which value is almost consistent with the simulation result of 21\% narrowing. 
These facts indicate that the sub-Planck nature is sufficiently observed for the even number of subtracted photons. Note that, in the case of three-photon subtraction, three negative regions should be in the Wigner function. However, in this experiment, due to the losses, the negative region only appears at the origin both in the experimental and simulated Wigner functions (Fig. \ref{fig.5}(h) and (l)).

Based on these results, the following improvements can be made to generate non-Gaussian states with more negativity. First, the WG-OPA should be shortened. The device used in this experiment is designed for cw light experiment and the waveguide length is relatively long (\SI{45}{mm}) to obtain sufficiently high parametric gain even at low pump power. When pumping with pulsed light as in this experiment, it is easy to achieve higher peak power. It is then possible to use a shorter waveguide which would allow keeping the loss in the WG-OPA lower than 5\%. Since the spatial mode of the output of the WG-OPA is close to $\text{TEM}_{00}$, the spatial mode match with the LO light from the fiber collimator is about 98\%, similar to previous studies \cite{Takase2022}. In contrast, there is still room for improvement in temporal mode matching. In this experiment, the second-order dispersion added by optical components such as long fibers is compensated by the WS, but it would be possible to improve temporal mode matching by correcting to higher-order dispersion and by optimizing the optical spectrum. In addition, the waveform of the pump pulse and the spectrum of the bandpass filter used before the TES, are known to affect the negativity of the Wigner function, so optimization of these parameters will improve the result \cite{Melalkia2022}.

The propagation loss of 5\% is mainly due to variable beam splitters consisting HWPs and PBSs for tapping and the HD. If these were replaced with fixed-ratio beamsplitters of high quality, the propagation loss could be reduced to 3\% \cite{Kawasaki2022}.

The photodiodes used in the HD have quantum efficiency of about 96\%. Very recently, photodiode with efficiency of 99\% at the telecommunication wavelength have been reported \cite{Shen2022}, which indicates the loss of the HD can be reduced in the near future. The effect of circuit noise can be effectively reduced by increasing the LO power. If the LO is pulsed light, however, the transimpedance amplifier of the HD will be saturated due to the high peak power, so no more LO light can be impinged in this experiment. This can be improved by designing a HD suitable for pulsed light experiment, and it can be possible to keep the loss from the HD to less than 1\%.

There is a relatively large loss on the idler. Many previous experiments have been conducted in the weak-pump limit, where the squeezing level is relatively small and only single-photon detection is performed. In the weak-pump limit, the efficiency of the idler has little effect on the purity of the generated state and is simply a matter of decreasing the event rate. However, in the case of multi-photon detection as in our experiment, it will be necessary to accurately detect two or more photons by PNRDs, and the loss on the idler becomes critical \cite{Provaznik2020}. Most of the loss on the idler in this experiment comes from the detection efficiency of the TES, which is caused by the surface coating and the coupling between the TES and the optical fiber. In the current proof-of-principle experiments, there are TESs with detection efficiencies approaching 95\% at \SI{1550}{nm} \cite{Lita2008}, so it is realistic to reduce the loss on the ider to less than 10\%.

If the above improvements can be made, the negative value of the Wigner function can be improved from the current value $-0.020$ to $-0.110$, which would be comparable to the experiments with cw light \cite{Takase2022}. Note that this estimation does not take into account improvements in temporal-mode shaping of the LO and spectral filtering on the idler. Furthermore, decreasing the squeezing level and the tap ratio would result in larger negative values at the cost of significantly reducing the count rate in multi-photon detection experiments for complex non-Gaussian state generation.

\section{Conclusion}
Non-Gaussian states with negative values of Wigner function without loss correction have been generated by subtracting up to three photons from pulsed squeezed light. 
The low-loss quasi-single-mode WG-OPA, the high-performance PNRD, and the phase-locked pulsed homodyne measurement system have led to the first observation of a negative value of the Wigner function in a ultra-short (8-ps) wave packet of light at the telecommunication wavelength (\SI{1545,32}{nm}). The generation rate limitation can be improved by using PNRDs with small timing jitter \cite{Lamas-Linares2013,Endo2021}, and further devising a heralded method \cite{Takase2021}. The results can be directly extended to the generation of more complicated non-Gaussian states including GKP qubits \cite{Gottesman2001,Su2019,Fukui2022,Takase2022-3}, which are the most promising candidate for fault-tolerant universal computer, and are a key technology to the realization of high-speed optical quantum computation.

\section*{Funding} 
Japan Science and Technology Agency (JPMJPR2254, JPMJMS2064, JPMJCR17N4); Japan Society for the Promotion of Science (18H05207, 20K15187, 22K20351); Ministerstvo Školství, Mládeže a Tělovýchovy (8C22001); Grantová Agentura České Republiky (22-08772S); Horizon 2020 Framework Programme (731473, 101017733, NONGAUSS (951737)).

\section*{Acknowledgments}
M.E. K. Takase, and W.A. were supported by the Research Foundation for Opto-Science and Technology. T.S. and K.Takahashi acknowledge financial supports from The Forefront Physics and Mathematics Program to Drive Transformation (FoPM). The authors acknowledge UTokyo Foundation and donations from Nichia Corporation.

\section*{Disclosures}
The authors declare no competing financial interests. 
\section*{Data availability}
Data underlying the results presented in this paper are not publicly available at this time but may be obtained from the authors upon reasonable request.

%%%%%%%%%%%%%%%%%%%%%%% References %%%%%%%%%%%%%%%%%%%%%%%%%

%%%%%%%%%% If using BibTeX:
\bibliography{photonsubt_v2}

\end{document}